\documentstyle{l-aa}

\begin{document}
\newcommand{\reff}{\noindent\hangindent=3em\hangafter=1}
\newcommand{\rb}{\right]}
\newcommand{\lb}{\left[}
\newcommand{\per}{$^{-1}$}
\newcommand{\mc}{\multicolumn}
\newcommand{\degree}{$^{\circ}$}
\newcommand{\SM}{M$_{\odot}$}
\newcommand{\citebare}[1]{{citename{#1} citeyear{#1}}}
\newcommand{\amucite}[1]{{(citename{#1} citeyear{#1})}}
 
\thesaurus{11(11.03.1; 11.03.4 Abell 85; 09.08.01 30 Dor; 13.19.2;
 13.25.2)}
 
\title{Observations of the $^{57}$Fe$^{+23}$ Hyperfine Transition in Clusters of Galaxies}
 
\author{
H.~Liang\inst{1,2,3}
\and J.~M.~Dickey\inst{2,4}
\and G.~Moorey\inst{2}
\and R.~D.~Ekers\inst{2}}
 
\offprints{H.~Liang, h.liang@bristol.ac.uk}
 
\institute{ H.H. Wills Physics Laboratory, The University of Bristol, Bristol BS8 1TL, UK
\and
 Australia Telescope National Facilities, CSIRO, Epping, NSW 2112, Australia
\and
 Institut d'Astrophysique Spatiale, Universit\'e Paris XI, 91405 Orsay
Cedex, France
\and
 University of Minnesota, 116 Church St. SE, Minneapolis, MN 55455 USA}

\date{Received ...}
 
\maketitle
 
\begin{abstract}
We present a search for the hyperfine transition of the
$^{57}$Fe$^{+23}$ ion at 3.071 mm in clusters of galaxies with the
ATNF Mopra telescope. The results are compared with a realistic
estimate of the peak brightness temperature of the line in a cooling
flow cluster A85, using the available X-ray data.

\end{abstract}
 
\input epsf

\section{Introduction}
It has been suggested by Sunyaev and Churazov (1984) that it might be
possible to detect the hyperfine transition of a Li-like
$^{57}$Fe$^{+23}$ ion line in clusters of galaxies. Hyperfine
transitions have been very important in astronomy. The first such
transitions was predicted for neutral hydrogen atoms (the famous 21cm
line) by van de Hulst in 1945 and detected in 1951. Atomic hyperfine
splitting is the result of the interaction between the magnetic moment
of the nucleus and the magnetic moment of the electron plus the
magnetic field generated by the circulating electrons. In the case of
the HI 21cm hyperfine transition at 1420 MHz, the ground state
$1^{2}S_{1/2}$ is split into 2 sublevels. Deuterium has a similar line
predicted at 327 MHz but there have been no detections to date despite
many efforts.  The spontaneous transition probability (or Einstein
coefficient) of the HI hyperfine transition is very low, $W=2.85\times
10^{-15}$ sec$^{-1}$. However, external excitation mechanisms such as
collisions with other H atoms and electrons or Ly$\alpha$ radiation
increase the transition rates enormously. Hyperfine splitting is not
present for all atoms or ions, since the interaction between the
nuclear magnetic moment and the electronic magnetic moment does not
occur if either of them is zero.  We know from quantum mechanics that
the total nuclear angular momentum (or nuclear spin) is zero for
nuclei with an even number of protons and neutrons, and that the total
electronic angular momentum of an ion with completely filled electron
shells is zero. Thus the commonly available isotope of iron $^{56}$Fe
has no hyperfine structure. However, Sunyaev and Churazov (1984; SC)
pointed out that the Li-like $^{57}$Fe$^{+23}$ ion exhibits hyperfine
structure with a higher spontaneous transition rate ($W=9.4\times
10^{-10}$ sec$^{-1}$) than that of the HI hyperfine transition
($W=2.85\times 10^{-15}$sec$^{-1}$), which compensates the low
abundance of the $^{57}$Fe isotope ($^{57}$Fe/$^{56}$Fe=2.3\%;
V\"okening 1989). The wavelength of the $^{57}$Fe$^{+23}$ hyperfine
transition has been calculated theoretically to be 0.3071(3)cm
(Shabaeva \& Shabaev 1992). The FWHM of the line is 40 MHz (or 130
km\,s$^{-1}$), assuming a cluster gas temperature of $2\times 10^{7}$K
and that the line width is due to thermal Doppler broadening alone.

Clusters of galaxies are known to possess hot gas of temperature
$10^{7}-10^{8}$K. This intracluster medium is found to be metal rich
with typically half solar abundance (Sarazin 1988). The early
X-ray observations detected the prominent ``7 Kev Fe line'' which
is a blend of Fe$^{+24}$ and Fe$^{+25}$ $K\alpha$ and $K\beta$
lines. The detection of this line not only provided the concrete proof
that the primary X-ray emission mechanism for clusters is thermal
bremstralung, but also showed that the intracluster gas has nearly
solar abundances and that at least some of the gas must have been
ejected from galaxies. Other Fe lines at lower energy such as
Fe$^{+23}$ L lines have been detected at $\sim 1$ keV in
clusters. Thus we know that Fe$^{+23}$ is abundant in cluster
environments. SC had estimated the
$^{57}$Fe$^{+23}$ line strength in the centre of the Perseus cluster
to be $\sim 1.5$ mK, assuming a central temperature of 1.4 keV inside a
10-20 kpc radius with an X-ray emission integral of $\sim 4.8\times
10^{66}$ cm$^{-3}$.

The detection of the $^{57}$Fe$^{+23}$ line will have significant
astrophysical implications on our understanding of the properties of
the galaxy clusters. Firstly, it will provide us with a direct measure
of the dynamics of the ICM, such as turbulence and accretion
flows. There have been great debates on the formation of tailed radio
sources in clusters of galaxies (e.g. Burns {\em et al.} 1986;
O'Donoghue {\em et al.} 1993; Roettiger {\em et al.} 1993), with some
suggesting that the sources are bent by their motions through the ICM
and others suggesting that the bending is the result of an intracluster
wind. This debate would benefit greatly from a direct measure of the
intracluster turbulence. Secondly, mm observations of the
$^{57}$Fe$^{+23}$ line will complement the X-ray measurements of the
distribution of iron in clusters and measure the $^{57}$Fe/$^{56}$Fe ratio
which would determine the nucleosynthesis history of the intracluster
gas. Finally, the mm line would provide a direct and high precision
measure of the cluster redshift since the spectral resolution at
mm-wavelength is much higher than in X-rays.

\section{Observations}

Our search for the $\lambda$ 3.1-mm $^{57}$Fe$^{+23}$ line was carried out in 
1996 September using the Mopra telescope of the Australia Telescope
National Facility.  The antenna is 22m in diameter of which 
the inner 15m is illuminated by the 3-mm feed, giving a beam 
width (FWHM) of 45'' and gain of 40 Jy K$^{-1}$.  The receiver
is a cooled SiS mixer with receiver temperature of 80 K which was
checked with hot-cold load measurements twice during the seven
day observation period.  Pointing was tested by using SiO masers,
the largest pointing error measured was 10''.  Atmospheric opacity
was checked by skydips, the zenith opacity varied between 0.06 and 0.12
for all but a few hours during the run. 

The spectrometer is an autocorrelator with two sections, each
covering 256 MHz with 256 channels (1 MHz per channel).  We overlapped these
two by 125 MHz so as to cover a total bandwidth of 384 MHz.  Roughly 
50 MHz on each side are compromised by the IF filter shape, leaving
a total bandwidth with consistent sensitivity of about 285 MHz which 
translates to 920 km s$^{-1}$ at a wavelength of 3.2 mm (the redshifted
wavelength).  Since the line width is not well known, we do not remove
baselines from the spectra, but simply edit the records on 
the basis of their raw rms noise.  

Each record consists of a pair of two minute on and off
source spectra from which we take the difference normalized by the
bandpass shape.  The reference position is offset by
six minutes of right ascension.  The expected spectral rms per 
record is about 20 mK, which is what we find in almost all cases.
A few records ($\sim 10\%$) show significantly higher noise due to baseline 
instabilities in the IF system.  They are removed from the average.
With this editing, the rms noise in the average decreases as integration
time $\sqrt{t_{int}}$ up to our maximum integration time of 10.5
hours.  Because the spectra have not been baseline fitted, there
remains some structure in the average spectra, but it is quite
broad band (typically 100 MHz wide or greater).  This baseline 
structure limits our ability to detect broad lines.

The clusters which we are searching for the $^{57}$Fe line are at
redshifts of $\simeq 0.06$.  Assuming rest frequency of $97.611
\pm 0.03$ GHz ($\lambda$ = 3.0713 mm) the cluster center frequencies
come out in the range 92 - 92.5 GHz.  More nearby clusters would not
necessarily give stronger signals, since the beam size of 45'' translates
to only 35 h$^{-1}$ kpc for a cluster at a redshift of 16,000 km s$^{-1}$.
Thus the emission region is probably considerably larger than the 
beam.  This is the reason for taking the reference position so far
from the source. 

\begin{figure}
\epsfxsize 250pt \epsfbox{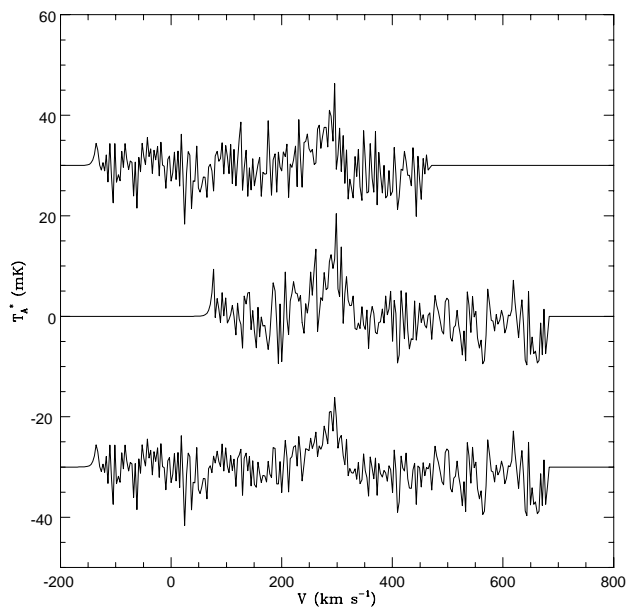}
\caption{Spectrum of the H41$\alpha$ line from 30 Dor. The antenna
temperature is plotted against the helio-centric velocity for the
data from both spectrometer channels separately and combined. The
vertical separations of 30 mK are for clarity in display and arbitary.}
\end{figure}
As tests of the sensitivity and stability of our system and 
observing strategy, we have observed several calibration lines.
The first test is simply to check the pointing using SiO masers
at 86 GHz.  We did these pointing checks every few hours for the 
first several days of the run observing five positions offset
by a half beamwidth from the tabulated source position. We always 
detected the masers within 10'' of their nominal positions and with 
their nominal line strengths.  Since both the beam size and the 
cluster sizes are much larger than this we have not introduced any
pointing offsets.  This test is limited in that the 86 GHz SiO 
frequency requires a different receiver tuning than the 92 GHz 
$^{57}$Fe line, so we checked the receiver performance by observing
test lines at 92 GHz.  The best candidates are $^{13}$CS line
from Orion (SiO position) and M17SW, whose frequency is 92.494 GHz
and the H41$\alpha$ recombination line at 91.955 GHz which we observed
in the 30 Dor region of the Large Magellanic Cloud.  The $^{13}$CS
line is so strong that it shows up in a single on-off pair, but the 
H41$\alpha$ line is weak enough to provide a good test of a moderately
long integration.  Our spectrum of this test line is shown in Figure 1;
the position chosen was the 20-cm continuum peak at 05:38:46.675,
-69:04:58.6 (J2000).  The total integration time is 112 minutes and 
the noise level is 2.5 mK, 2.0 mK would be predicted by the radiometer 
equation, so the system is evidently performing well for integration
times of an hour or so.

\begin{table}
\caption[]{Summary of Observations}
{\small
\begin{tabular}{llllll}\hline
\multicolumn{1}{l}{Name} & \multicolumn{1}{l}{$\alpha$ (2000.0)} &
 \multicolumn{1}{l}{$\delta$ (2000.0)} &
\multicolumn{1}{l}{$t_{int}$} & \multicolumn{1}{l}{$T_{b}$} & \multicolumn{1}{l}{$cz$ range} \\ \noalign{\smallskip}
\multicolumn{1}{l}{} & \multicolumn{1}{l}{~h~~m~~~s} &
 \multicolumn{1}{l}{~~~$^\circ$~~~$'$~~~$''$} &
\multicolumn{1}{l}{min.} & \multicolumn{1}{l}{mK} & \multicolumn{1}{l}{$10^{2}$ km~s$^{-1}$}\\ 
\hline
A85  & 00 41 50.9  & $-$09 18 10.7   & 630  & 5  & 160-168 \\
A3266 & 04 30 33.4  & $-$61 33 34.3  & 108 &  21 & 174-180 \\
A3391 & 06 25 16.0 & $-$53 39 39.0  & 124 &  20 & 160-168 \\
A3667 & 20 12 27.4 & $-$56 49 35.7  & 282 &  18 & 160-168 \\
\hline
\end{tabular}}
\end{table}

Table 1 lists the clusters which we have searched for $^{57}$Fe emission.
Columns 1-3 give the cluster name and position, column 4 gives the
integration time on-source, column 5 gives the detection threshold, and
column 6 gives the useful velocity range.  The detection threshold 
is a strong function of the width of the line because we have 
subtracted no baselines from our spectra.  The threshold given on 
Table 1 applies to lines narrower than about 150 km s$^{-1}$ in
full-width.  For broader lines the residual baselines in the 
receiver system begin to degrade our detection threshold.  This 
effect is illustrated in Figure 2, which is the pair of spectra 
for the cluster A85 for the two spectrometer channels.
\begin{figure}
\epsfxsize 250pt \epsfbox{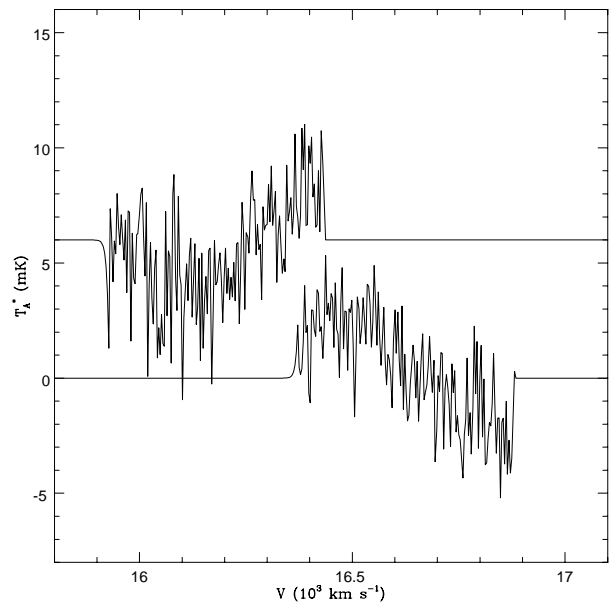}
\caption{Spectrum for the cluster A85. Antenna temperature is plotted
against the helio-centric velocity for the 2 spectrometer
channels separately. The vertical separation of 6 mK is arbitary, for
clarity in display.}
\end{figure}
The vertical 
offset between the two is arbitrary, here it has been set at 6 mK
for convenience.  One channel is dominated by a linear baseline,
while the other shows primarily a quadratic baseline shape.  If we
were to remove first and second order baselines and align the 
two channels we would obtain a spectrum with no features above 
2 mK or less (depending on the smoothing).  On the other hand, if
cluster lines are as broad as several hundred km s$^{-1}$ then 
this baseline removal could remove a real feature, and so our
threshold would have to be raised; we cannot rule out the lump
between 16200 and 16500 km/s being part of a real feature of
strength 3 mK and linewidth $\simeq$300 km s$^{-1}$.
For linewidths of 200 km/s or greater we must set the threshold at 5 mK
in this case.  For the other clusters on table 1 there are similar
baseline irregularities that set the maximum linewidth corresponding
to the threshold at about 150 km/s.

\section{Discussions and Conclusion}
Following the formalism given in Sunyaev and Churazov (SC;1984), we can
calculate the expected line intensity for a cluster given the cluster gas
parameters derived from X-ray observations.

The peak brightness temperature of the 
$^{57}$Fe$^{+23}$ hyperfine transition should be
\begin{equation}
\Delta T_{b}=\frac{c^{2}}{8\pi^{3/2}\nu} \frac{h}{k} \frac{1}{\Delta \nu_{D}} \alpha \frac{g_{u}}{g_{l}} \Phi
\end{equation}
where $g_{u}$,$g_{l}$ are the statistical weights of the hyperfine
sublevels, $\Delta \nu_{D}$ is the Doppler line width (i.e. the
half-width at $e^{-1}$ of the maximum), $\alpha$ is the cluster
isotope abundance relative to hydrogen and the function $\Phi$ is
given by
\begin{equation}
\label{eq:phi1}
\Phi= \int{\frac{W\delta(T)\frac{1}{2} n^{2}_{e}\langle\sigma v\rangle}{W(1+N)+\frac{1}{2}n_{e}\langle\sigma v\rangle+\frac{g_{u}}{g_{l}}WN+\frac{g_{u}}{g_{l}}\frac{1}{2}n_{e}\langle\sigma v\rangle} dl}
\end{equation}
where $W$ is the spontaneous transition probability, $\delta(T)$ is
the fraction of ions of a given species at temperature T, $N$ is the
occupation number of the microwave background at the wavelength of
transition, $\langle\sigma v\rangle$ is the excitation coefficient by
resonance scattering which is the dominant excitation process (SC). Note
that Eq.~(\ref{eq:phi1}) differs slightly from equation~(2) in SC (see
Appendix).  

In the case of clusters of galaxies where $WN\gg
\frac{1}{2}n_{e}^{2}\langle \sigma v\rangle$ (or $n_{e} \ll 1$
cm$^{-3}$), Eq.~(2) is reduced to
\begin{equation}
\Phi=\int \frac{\delta(T) \frac{1}{2} \langle\sigma
v\rangle}{1+N+\frac{g_{u}}{g_{l}}N} n_{e}^{2} dl
\end{equation}
Assuming a constant temperature, we have from Eq.~(1) and Eq.~(3)
\begin{equation}
\Delta T_{b} \propto \int n_{e}^{2} dl \propto S_{x}
\end{equation}
where $S_{x}$ is the X-ray surface brightness.  If we assume solar
abundance for clusters, where solar $^{57}$Fe abundance was taken to
be $8.8\times10^{-7}$ (SC), and isothermal cluster gas temperature
within the Mopra beam to be at its optimum value of $2\times 10^{7}$K
where $\delta(T)$ has its peak value of $\sim 0.3$ (Jacobs {\em et
al.}  1977), we can deduce a simple relation between the peak brightness
temperature of the $^{57}$Fe line and the X-ray surface brightness:
\begin{equation}
\Delta T_{b} \sim 100 S_{x}
\end{equation}
where $\Delta T_{b}$ is the peak brightness temperature of the line in
units of mK before convolution with the beam of the radio telescope,
and $S_{x}$ is the X-ray surface brightness in units of
ergs\,s\per\,cm$^{-2}$\,ster\per in the ROSAT band of [0.1,2.4] keV
after corrections for the telescope response and Galactic absorption.
In the case of an average cluster of galaxies, where $n_{e}\sim
10^{-3} - 10^{-2}$ cm$^{-3}$, the peak brightness temperature would be
$\sim 0.001$ mK. However in the centre of cooling flow clusters of
galaxies where the central density could be as high as $n_{e}\sim 0.1
- 0.2$ cm$^{-3}$, the peak brightness temperature could only reach $\sim
0.1$ mK in the most extreme cases, contrary to the estimate of $\sim
1.5$ mK given by SC for the Perseus cluster.

The cluster A85 is considered a classic example of a cooling flow
cluster with high X-ray luminosity and thus one of the best candidates
for this experiment.  To obtain a realistic prediction for the
$^{57}$Fe line brightness for A85, we examine the X-ray data available
for the cluster.  The ROSAT PSPC data for A85 has been analysed by
Pislar {\em et al.}  (1996), who found an X-ray luminosity of
$9.3\pm0.2 \times 10^{44}$ ergs s$^{-1}$ within a radius of 1.4 Mpc in
the ROSAT band of [0.1-2.4]keV. The X-ray
surface brightness was fitted with a $\beta$ model given by,
\begin{equation}
S_{x}\propto [1+(r/r_{c})^2]^{-3\beta+1/2}
\end{equation}
which corresponds to 
\begin{equation}
n_{e}(r)=n_{0} [1+(r/r_{c})^2]^{-3\beta/2}
\end{equation}
for the electron density distribution, where $n_{0}=0.01$cm$^{-3}$,
$\beta=0.438$ and $r_{c}=32$kpc ($25''$).  However, the surface
brightness profile obtained with the ROSAT HRI (Prestwich {\em et al.}
1995) which has a resolution of $5''$ (c.f. $20''$ the PSPC) clearly
shows that there is an excess above the $\beta$ model within the
central $20''$ radius. Taking the X-ray surface brightness from HRI,
we obtain from the above relation a peak brightness temperature of
0.003 mK for the $^{57}$Fe line. This predicted line brightness can be
increased, if the abundance of $^{57}$Fe is significantly different
from the terrestrial abundance and/or the distribution of Fe increases
towards the centre.  However, such effects are unlikely to increase
the detectability of the line by more than a factor of 10.

This analysis shows that the expected brightness temperature of the
$^{57}$Fe line to be $\ll 1$ mK in cluster cores, even for an archetype
cooling flow cluster such as Abell 85. We have shown that the MOPRA
telescope can reach a sensitivity level comparable to the original
prediction given by SC of the line intensity but is certainly inadequate
given the more realistic estimates shown above.

\begin{acknowledgements}  We would like to thank Malcom Gray, Mike Masheder and Rashid Sunyaev for useful discussions, Graeme Carrad for helping with the instrument set up. HL would like to thank the ATNF and INSU (Franco-Australian collaboration fund) for support.
\end{acknowledgements}
 
\appendix
\section*{Appendix: derivation of Eq.~(1) \& (2)}
The number
of transitions upwards equals to the number of transitions downwards:
\begin{equation}
n_{u}W + n_{u}B_{ul}B_{\nu}+n_{u}c_{ul} = n_{l}B_{lu}B_{\nu}+n_{l}c_{lu}
\end{equation}
where $n_{u}$, $n_{l}$ are the number in the upper and lower states
respectively, $W$ and $B$ are the Einstein A and B coefficients relating
to spontaneous and stimulated emission/absorption, $B_{\nu}$ is the
intensity of the microwave background and $c$ is the collisional
excitation/de-excitation rate. The $W$,$B$ and $c$ coefficients are
related as follows (Rybiki and Lightman 1979):
\begin{equation}
B_{lu}=\frac{g_{u}}{g_{l}}B_{ul},\; B_{ul}B_{\nu}=WN,\; c_{lu}=\frac{g_{u}}{g_{l}}c_{ul}\exp(-h\nu/kT)
\end{equation}
where $N=(\exp(h\nu/kT)-1)^{-1}$ is the occupation number of the
microwave background. As was discussed in SC, the dominant excitation
mechanism in this case is resonance scattering of an electron by the
ion and the rate of excitation is given by
$c_{ul}=\frac{1}{2}n_{e}\langle\sigma v\rangle$, where $\langle\sigma
v\rangle$ is the average of the product of the resonance scattering
cross section and the relative velocity between the electron and the
ion.

The peak intensity of the line is given by
\begin{equation}
I_{0}=(N_{u} W+N_{u}B_{ul}B_{\nu}-N_{l}B_{lu}B_{\nu}) \frac{h\nu_{0}}{4\pi} \phi(\nu_{0})
\end{equation}
where $N_{u}$,$N_{l}$ are the total number of ions in the upper and
lower levels per unit area along the line of sight respectively, $W$
is the spontaneous emission coefficient or Einstein A-coefficient,
$\nu_{0}$ is the central frequency of the line and $\phi$ is the line
profile. Assuming thermal doppler broadening alone, we have
\begin{equation}
\phi(\nu)=\frac{1}{\sqrt{\pi}\Delta\nu_{D}}e^{-(\nu-\nu_{0})^{2}/\Delta\nu^{2}_{D}}
\end{equation}
where $\Delta\nu_{D}=\sqrt{\frac{2kT}{Mc^{2}}}\nu_{0}$ is the doppler
width which is related to the FWHM $\Delta\nu_{L}=2\sqrt{\ln{2}}\Delta\nu_{D}$.  
In our case where $h\nu\ll kT$, we have the peak brightness
temperature of the line given by $\Delta
T_{b}=\frac{c^{2}}{2k\nu^{2}}I_{0}(\nu_{0})$. By substituting
Eq.~1 into the expression for $I_{0}$ in Eq.~(3), we have
\begin{equation}
\Delta T_{b}=\frac{c^{2}}{8\pi^{3/2}\nu}\frac{1}{\Delta\nu_{D}}\frac{h}{k} 
(N_{l}\frac{g_u}{g_{l}}\frac{1}{2}n_{e}\langle\sigma v\rangle
-N_{u}\frac{1}{2}n_{e}\langle\sigma v\rangle)
\end{equation}
where 
\begin{equation}
N_{u}= \int \alpha \delta(T) \frac{n_{u}}{n_{l}+n_{u}} \frac{n_{H}}{n_{e}} n_{e} dl
\end{equation}
with $ \frac{n_{H}}{n_{e}} \sim 1$ and $ \alpha \delta(T)$ giving the abundance of $^{57}$Fe$^{+23}$ in the cluster, and $N_{l}=(n_{l}/n_{u})N_{u}$.
From Eq.~(1) and (2), we have 
\begin{equation}
\frac{n_{u}}{n_{l}}=\frac{g_{u}}{g_{l}}\frac{WN+\frac{1}{2}n_{e}\langle\sigma v\rangle}{W(N+1)+\frac{1}{2}n_{e}\langle\sigma v\rangle}
\end{equation}

Thus the peak brightness temperature is given by
\begin{equation}
\Delta T_{b}=\frac{c^{2}}{8\pi^{3/2}\nu} \frac{h}{k} \frac{1}{\Delta \nu_{D}} \alpha \frac{g_{u}}{g_{l}} \Phi
\end{equation}
where 
\begin{equation}
\Phi= \int{\frac{W\delta(T)\frac{1}{2} n^{2}_{e}\langle\sigma v\rangle}{W(1+N)+\frac{1}{2}n_{e}\langle\sigma v\rangle+\frac{g_{u}}{g_{l}}WN+\frac{g_{u}}{g_{l}}\frac{1}{2}n_{e}\langle\sigma v\rangle} dl}
\end{equation}
The term $\frac{g_{u}}{g_{l}}\frac{1}{2}n_{e}\langle\sigma v\rangle$
in the denominator is absent in equation~2 of SC, however in cases where 
$WN\gg\frac{1}{2}n_{e}^{2}\langle \sigma v\rangle$ this term is insignificant.

\end{document}